\documentclass[prb,aps]{revtex4}
\usepackage{graphicx}
\usepackage{epsf}
\newcommand{\e}{{\rm e}}
\newcommand{\ep}{\varepsilon}
\newcommand{\dt}{\Delta t}
\newcommand{\cat}{{\cal T}}

\newcommand{\bea}{\begin{eqnarray}}
\newcommand{\eea}{\end{eqnarray}}
\newcommand{\be}{\begin{equation}}
\newcommand{\ee}{\end{equation}}
\newcommand{\ba}{\begin{eqnarray}}
\newcommand{\ea}{\end{eqnarray}}

\newcommand{\nn}{\nonumber}
\newcommand{\la}{\label} 
 
\newcommand{\hT}{\hat{T}}
\newcommand{\hV}{\hat{V}}
\newcommand{\hH}{\hat{H}}
\newcommand{\htH}{\hat{H}_A}
\newcommand{\pa}{\partial}

\def\t1{e_{_T}}
\def\v1{e_{_V}}
\def\tvf{e_{_{TV}}}
\def\ct{e_{_{TTV}}}
\def\cv{e_{_{VTV}}}

\def\fft#1#2{{#1 \over #2}}

\begin{document}
\title{Explicit symplectic integrators for solving non-separable Hamiltonians.}

\author{Siu A. Chin}

\affiliation{Department of Physics, Texas A\&M University,
College Station, TX 77843, USA}

\begin{abstract}

By exploiting the error functions of explicit symplectic integrators
for solving separable Hamiltonians, I show that it is possible to
develop explicit, time-reversible symplectic integrators for solving  
non-separable Hamiltonians of the product form. The algorithms are 
unusual in that they are of fractional orders. 

\end{abstract}
\maketitle

\section {Introduction}
Symplectic integrators\cite{yoshi,mat92,mcl02,hairer} are the
methods of choice for solving diverse physical problems in
classical\cite{yoshi,skeel,chinchen03,chinsante}, 
quantum\cite{shen,serna,chinchen01,chinchen02,fchinm,auer,ochin,chinkro}, 
and statistical\cite{ti,li,jang,chincor} mechanics. 
For separable Hamiltonians, the problem is well understood and many explicit 
integrators are available\cite{yoshi,mat92,mcl02,hairer}. However, for
non-separable Hamiltonians, only implicit algorithms are 
known\cite{yoshi,mat92,mcl02,hairer}. 
It is generally believed that no explicit algorithms can be developed for solving
non-separable Hamiltonians\cite{yoshi,mat92}. In this work, I show that 
this is not the case. Explicit, time-reversible algorithms can be developed 
to solve a selected class of non-separable Hamiltonians. The idea is to model 
non-separable Hamiltonians by the error terms of explicit algorithms when 
solving separable Hamiltonians. By a suitable choice of factorization (or split) 
coefficients, the explicit algorithm can be made to solve error 
Hamiltonians which are generally non-separable.

In the usual study of symplectic integratiors, one seeks to eliminate 
error terms in order to produce higher order algorithms. These error 
terms are therefore not of direct interest and are rarely studied in 
their own right. In this work, these error terms are the non-separable 
Hamiltonians we seek to solve. The method can solve non-separable 
Hamiltonians of the product form, (sum over repeated indices)
\be
H=T_i({\bf p})V_{ij}({\bf q})T_i({\bf p}),
\la{liop}
\ee
provided that 
\be
T_i({\bf p})=\frac{\partial}{\partial p_i} T({\bf p}).
\ee
and
\be
V_{ij}({\bf q})=\frac{\partial^2}{\partial q_i\partial q_j} V({\bf q}).
\ee
For one degree of freedom, given $T^\prime(p)$ and $V^{\prime\prime}(q)$, 
$T(p)$ and $V(q)$ can always be obtained by integration. 

In the next section we will briefly summarize essential aspects of
symplectic integrators and their error functions, followed by
our explicit integrator for solving the above non-separable Hamiltonian.
Higher order algorithms are discussed in Section IV. 

\section {Symplectic integrators }
Given a dynamical variable $W(q_i,p_i)$ and	a
Hamiltonian function $H(q_i,p_i)$, the former is evolved by
the later via the Poisson bracket, and therefore by the
corresponding Lie operator\cite{dragt} $\hat H$ associated with the
function $H(q_i,p_i)$,
\ba
\fft{dW}{dt} &=&\{W,H\}\nn\\
             &=& \Bigl(\fft{\pa H}{\pa p_i}\fft{\pa}{\pa q_i}
         -\fft{\pa H}{\pa q_i}\fft{\pa}{\pa p_i}\Bigr)W
             = \hH W, 
\la{liou}
\ea
via exponentiation,
\be
W(t+\ep)={\rm e}^{\ep \hH}W(t).
\la{formsol}
\ee
For a separable Hamiltonian,
\be
H({\bf q},{\bf p}) = T({\bf p})+V({\bf q}),
\la{seph}
\ee
the corresponding Hamiltonian operator is also separable,
$
\hH=\hT+\hV , \la{htv}
$
with $\hat{T}$ and $\hat{V}$ given by
\be
\hT \equiv\{\cdot,T\}=\fft{\pa T}{\pa p_i}\fft{\pa}{\pa q_i},
 \la{htop}
 \ee
 \be
\hV \equiv\{\cdot,V\}= -\fft{\pa V}{\pa q_i} \fft{\pa}{\pa p_i}.
\la{hvop}
\ee
Their corresponding evolution operators
${\rm e}^{\,\ep\, \hT}$ and ${\rm e}^{\,\ep\, \hV}$ 
then shift $q_i$ and $p_i$ forward in time via
\ba
&&q_i(\ep)={\rm e}^{\,\ep\, \hT}q_i= q_i+\ep\,\fft{\pa T}{\pa p_i},\nn\\
&&p_i(\ep)={\rm e}^{\,\ep\, \hV}p_i=p_i-\ep\,\fft{\pa V}{\pa q_i} .
\la{pqsh}
\ea
Conventional symplectic integrators correspond to approximating 
the short time evolution operator  ${\rm e}^{\ep \hH}$ in the product form
\be
{\rm e}^{\ep(\hT+\hV)}\approx\prod_{i=1}^N
{\rm e}^{t_i\ep \hT}{\rm e}^{v_i\ep \hV}, \la{prod}
\ee
resulting in an ordered sequence of displacements (\ref{pqsh}) which defines
the resulting algorithm. Here, we will consider only time-reversible,
symmetric factorization schemes 
such that  either $t_1=0$ and
$v_i=v_{N-i+1}$, $t_{i+1}=t_{N-i+1}$, or $v_N=0$ and
$v_i=v_{N-i}$, $t_{i}=t_{N-i+1}$.  

The product of operators in (\ref{prod}) can be combined by use of the
Baker-Campbell-Hausdorff (BCH) formula to give
\be
\prod_{i=1}^N
{\rm e}^{t_i\ep \hT}{\rm e}^{v_i\ep \hV}
={\rm e}^{\ep\htH},
\la{tbopt}
\ee
where the approximate Hamiltonian operator $\htH$ 
has the general form 
\ba
&&\htH = \t1\hT+\v1\hV+\ep^2\ct[\hT\hT\hV]\nn\\
&&\qquad\quad+\ep^2\cv[\hV\hT\hV] +O(\ep^4)
\la{hopbk}
\ea
where $\t1$, $\tvf$, $\ct$, etc., are functions of $\{t_i\}$ and $\{v_i\}$ and
where condensed commutator brackets, $[\hT\hT\hV]=[\hT,[\hT,\hV]]$, 
$[\hT\hV\hT\hV]=[\hT,[\hV,[\hT,\hV]]]$, etc., are used. 
From the way Lie operators are defined via (\ref{liou}), one can
convert operators back to functions\cite{yoshi,chinsante} 
via $[T,V]\rightarrow\{V,T\}=-\{T,V\}$, yielding
\ba
&& H_A = \t1 T+\v1 V+\ep^2\ct\{TTV\}\nn\\
&&\quad\quad\quad+\ep^2\cv\{VTV\} +O(\ep^4),
\la{hopft}
\ea
where again, condensed Poisson brackets, $\{TTV\}=\{T,\{T,V\}\}$, etc., are used. 
For a separable Hamiltonian of the
form (\ref{seph}), we have 
\bea
\{TV\}&=&-\fft{\pa T}{\pa p_j}\fft{\pa V}{\pa q_j}=-T_jV_j\, ,\nn\\
\{TTV\}&=&-\fft{\pa T}{\pa p_i}\fft{\pa \{T,V\}}{\pa q_i}
=T_iV_{ij}T_j\, ,\la{httv}\\
\{VTV\}&=&\fft{\pa V}{\pa q_i}\fft{\pa \{T,V\}}{\pa p_i}=-V_iT_{ij}V_j.
\la{hvtv}
\eea
By choosing $\{t_i\}$ and $\{v_i\}$ such that
\be
 \t1=\v1=0,
\la{tvzero}
\ee
and either $\cv=0$, or $\ct=0$,
the algorithm would then be solving the non-separable Hamiltonian,
either
\be
H_{TTV}=T_iV_{ij}T_j\qquad {\rm or}\qquad H_{VVT}=V_iT_{ij}V_j.
\ee 

\section {Solving non-separable hamiltonians}

The following
factorization scheme gives,
\ba
{\cal T}(\ep)&\equiv&
{\rm e}^{\ep v_2\hV}
{\rm e}^{\ep t_2\hT}
{\rm e}^{\ep v_1\hV}
{\rm e}^{\ep t_1\hT}
{\rm e}^{\ep v_0\hV}
{\rm e}^{\ep t_1\hT}
{\rm e}^{\ep v_1\hV}
{\rm e}^{\ep t_2\hT}
{\rm e}^{\ep v_2\hV}\nn\\
&=&\exp(\,\ep^3[\hT\hT\hV]+\ep^5 E_5+\ep^7 E_7+\ep^9 E_9\cdots),
\la{expalg}
\ea
with $v_0=-2(v_1+v_2)$, $t_1=-t_2$, $v_2=-v_1/2$ and $v_1=1/t_2^2$. There is
one free parameter $t_2$ that one can choose to minimize the resulting error, but
not be set to zero. As examplfied by (\ref{httv}) and (\ref{hvtv}), for a separable
Hamiltonian $H=T+V$, higher order brackets of the form $\{T,Q\}$, $\{V,Q\}$
have opposite signs. Thus one should choose algorithms with $e_{TQ}=e_{VQ}$ to maximize
error cancellations\cite{chincor}. This is the basis for symplectic 
corrector\cite{wis96} 
or processed\cite{mar97,blan99} algorithms.
The choice of $t_2=-6^{1/3}\approx -1.82$ forces $e_{TTTTV}=e_{VTTTV}$ and 
would be a good starting value.
The RHS of (\ref{expalg}) is the evolution operator for the non-separable
Hamiltonian $H_{TTV}$ with time step $\Delta t=\ep^3$ and
leading error terms $O(\ep^5)$. Thus the parameter $\ep$ used by the
integrator is $\ep=\root 3\of{\Delta t}$. Since $\ep^5=\Delta t^{5/3}$,
the basic algorithm (\ref{expalg}) in terms of $\dt$ reads,
\be
{\cal T}(\dt)
=\exp\dt\left( [\hT\hT\hV]+\dt^{2/3} E_5+\dt^{4/3} E_7+\dt^{6/3} E_9\cdots\right).
\la{torder}
\ee
The order of the algorithm ${\cal T}(\dt)$ (the leading error in the Hamiltonian) is 
therefore only 2/3. We will discuss this and higher order algorithms in the
next section.

By interchange $\hT\leftrightarrow\hV$ everywhere, but
keeping the coefficents intact, the RHS of (\ref{expalg}) goes over to
\be
 \e^{\ep^3 [\hT\hT\hV]}\rightarrow\e^{\ep^3[\hV\hV\hT]},
\ee
and the basica algorithm ${\cal T}(\dt)$ solves the non-separable Hamiltonian $H_{VVT}$.
In both cases, the final force or velocity can be
re-used at the start of the next iteration. Thus both algorithms
require four-force and four-velocity evaluations. 

For one degree of freedom, any Hamiltonian of the form
\be
H=f(p)g(q)
\la{hprod}
\ee
can be solved. To test the algorithm, we solve the non-separable Hamiltonian
\be
H_{TTV}=(1+\frac{p^2}2)^2(1+q^2),
\la{h1d}
\ee
where the phase trajectory is harmonic near the origin,
but highly distorted at larger values of $(p,q)$. The
algorithm's separable Hamiltonian is
\be
H=p+\frac16 p^3+\frac12 q^2+\frac1{12}q^4.
\la{algham}
\ee

In Fig.\ref{fig1} we compare the phase trajectories produced by algorithm
(\ref{expalg}) with exact
trajectories deduced from (\ref{h1d}). We set
$t_2=-2$ and use a relatively large value of $\Delta t=0.005$
so that discrepances can be seen. The four trajectories are
started at $p_0=0$ and $q_0=0.5$, 1.0, 1.5, and 2.0 respectively.
The error is largest at the positive maximum of $p$ and next largest at 				 
the negative maximum of $p$. 
In each case, the error can be further reduced by making 
$t_2$ more negative than -2. We did not bother with this refinement here,
but this will be important in the 2D case discussed below.

We will demonstrate that $\cat(\dt)$ indeed converges as
$\Delta t^{2/3}$ in the next section.
      
For more than one degree of freedom, the generalization of (\ref{hprod})
to
\be
H=\sum_i f_i(p_i)g_i(q_i)
\ee
can always be solved. However, it is more interesting to generalize
(\ref{algham}) to N-dimension by reinterpreting $p$ and $q$ as
radial coordinates: $p=\sqrt{\sum_i p_i^2}$, $q=\sqrt{\sum_i q_i^2}$.
For any radial potential $V(q)$,  
\be
V_{ij}=\frac{V^\prime}q\delta_{ij}
+\Bigl( V^{\prime\prime}-\frac{V^\prime}q\Bigr) \hat q_i\hat q_j,
\ee
where here $\hat {\bf q}$ is the unit vector. Thus the non-separable
Hamiltonian $H_{TTV}$ corresponding
to the radial Hamiltonian (\ref{algham}) is
\be
H_{TTV}=(1+\frac{p^2}2)^2
\Bigl[
1+\frac13 q^2+\frac23 q^2(\hat{\bf p}\cdot\hat{\bf q})^2.
\Bigr]
\la{hnd}
\ee
This can again be solved by our explicit
integrator (\ref{expalg}). In two-dimension, most trajectories are
not closed and are likely to be chaotic. However, for some special initial
configurations, a rich variety of closed orbits can be found. Fig.\ref{fig2} 
shows a sample of three such closed orbits. For this calculation, since
the order of the algorithm is only $2/3$, reducing the step size is not
efficient in achieving higher accuaracy. Instead, we find that 
the error can be substantially reduced by changing
$t_2$ to $\approx -3$. For the circle, triangle and the twisted orbits of
Fig. 3, the step sizes used were,
$\Delta t=0.0012$, 0.001, and 0.0005 respectively.

Finally, the standard kinetic energy term 
\be
T({\bf p})=\frac12 p_ip_i
\ee
produces
\bea
H_{TTV}=\{TTV\}&=&p_{i}V_{ij}p_{j}\, ,\la{erttv}\\
H_{VTV}=\{VTV\}&=&-V_{i}V_{i},
\la{errt}
\eea
and only $H_{TTV}$ is non-separable. Here, $V_{ij}$ can be
viewed as a position-dependent inverse mass matrix. This work
shows that if $V_{ij}$ can be derived from a potential function
$V({\bf q})$, then this non-separable Hamiltonian can also be solved
by our explicit algorithm. Also, by itself, this quadratic
Hamiltonian does not possess closed orbits for most $V({\bf q})$,
thus explaining why this error term would disrupt closed orbit 
of the original Hamiltonian at large $\ep$.

\section{Higher order algorithms}

In the previous section, we have shown that the primitive
algorithm ${\cal T}(\dt)$ does work and reproduces the correct
phase trajectory. However, its 2/3-order convergence is
very poor and requires extremely small $\dt$ to produce accurate
results. To demonstrate its fractional order convergence, we return
to the one-dimensional case (\ref{h1d}) and integrate from
$t=0$, $p_0=0$, $q_0=2$ to $t=T_{1/4}\equiv 0.385841$, $p(t)=-1.569196$, $q(t)=0$,
corresponding to a quarter, clockwise rotation of the outermost
phase trajectory of Fig.{\ref{fig1}. In Fig.\ref{fig3}, the relative
error of the Hamiltonian (\ref{h1d}) at $t=T_{1/4}$ is plotted as
a function of $\dt$. The error of ${\cal T}(\dt)$ can be perfectly
fitted with the power law $-2\dt^{2/3}$, but due to
this fractional power, the convergence at small $\dt$ is very poor.
Fortunately, the error structure (\ref{torder}) of ${\cal T}(\dt)$ allows
simple ways of generating higher order symplectic algorithms. The
triplet-construction of Creutz and Gocksch\cite{cre89} and
Yoshida\cite{yos90} can produce arbitrary high order algorithms
such as the following 4/3rd order algorithm
\be
\cat_{4/3}(\dt)=\cat\left(\frac{\dt}{2-s}\right)\cat\left(-\frac{s\dt}{2-s}\right)
\cat\left(\frac{\dt}{2-s}\right)
\la{sym43}
\ee
with $s=2^{3/5}$ and the following 6/3rd=2nd-order algorithm   
\be
\cat_{2}(\dt)=\cat_{4/3}\left(\frac{\dt}{2-s}\right)
\cat_{4/3}\left(-\frac{s\dt}{2-s}\right)\cat_{4/3}\left(\frac{\dt}{2-s}\right)
\la{sym63}
\ee
with $s=2^{3/7}$. As can be seen in Fig.{\ref{fig3},
these higher order symplectic algorithms are orders of magnitude better than
the basic algorithm $\cat(\dt)$. 
The disadvantage of the triplet-construction is that the computational effort triples in going 
from order $k/3$ to $(k+2)/3$. For example, the second-order algorithm
$\cat_{2}(\dt)$ requires three evaluations of $\cat_{4/3}(\dt)$, or
nine evaluations of $\cat(\dt)$. Alternatively, arbitrary high order
algorithms can also be obtained via the Multi-Product Expansion(MPE)\cite{chin084},
with only quadratically growing computational efforts. For example, by
replacing $k^2_i\rightarrow k_i^{2/3}$ in \cite{chin084}, one obtains 
\be
\cat^{MPE}_{4/3}(\dt)=\frac{1^n}{1^n-2^n}\cat(\dt)
+\frac{2^n}{2^n-1^n}\cat^2\left(\frac{\dt}{2}\right)
\la{mpe43}
\ee
\ba
&&\cat^{MPE}_{2}(\dt)=\frac{(1^n)^2}{(1^n-2^n)(1^n-3^n)}\cat(\dt)
+\frac{(2^n)^2}{(2^n-1^n)(2^n-3^n)}\cat^2\left(\frac{\dt}{2}\right)\nn\\
&&\qquad\qquad\qquad\qquad\qquad\qquad\qquad\quad\qquad\ 
+\frac{(3^n)^2}{(3^n-1^n)(3^n-2^n)}\cat^3\left(\frac{\dt}{3}\right)
\la{mpe63}
\ea
with $n=2/3$ in both cases. Here, $\cat^{MPE}_{2}(\dt)$ only requires six evaluations of
$\cat(\dt)$. The disadvantage of MPE is that it is no longer symplectic, but
is like Runge-Kutta-Nystr\"om type algorithms. However, as shown in Fig.\ref{fig3},
their energy error can be much smaller than the triplet symplectic algorithms.

\section{Concluding summary}
In this work, we have shown that explicit symplectic integrators 
can be devised to solve a selected class of non-separable Hamiltonians. 
Any non-separable Hamiltonian which can be modelled by the error terms 
of an explicit integrator can be solved by the same integrator with changed
splitting coefficients. The initial explicit algorithm is only of fractional
order $\Delta t^{2/3}$, but higher order algorithms can be easily
obtained by use of the triplet construction or the multi-product expansion.


\newpage
\begin{figure}
	\vspace{0.5truein}
	\centerline{\includegraphics[width=0.8\linewidth]{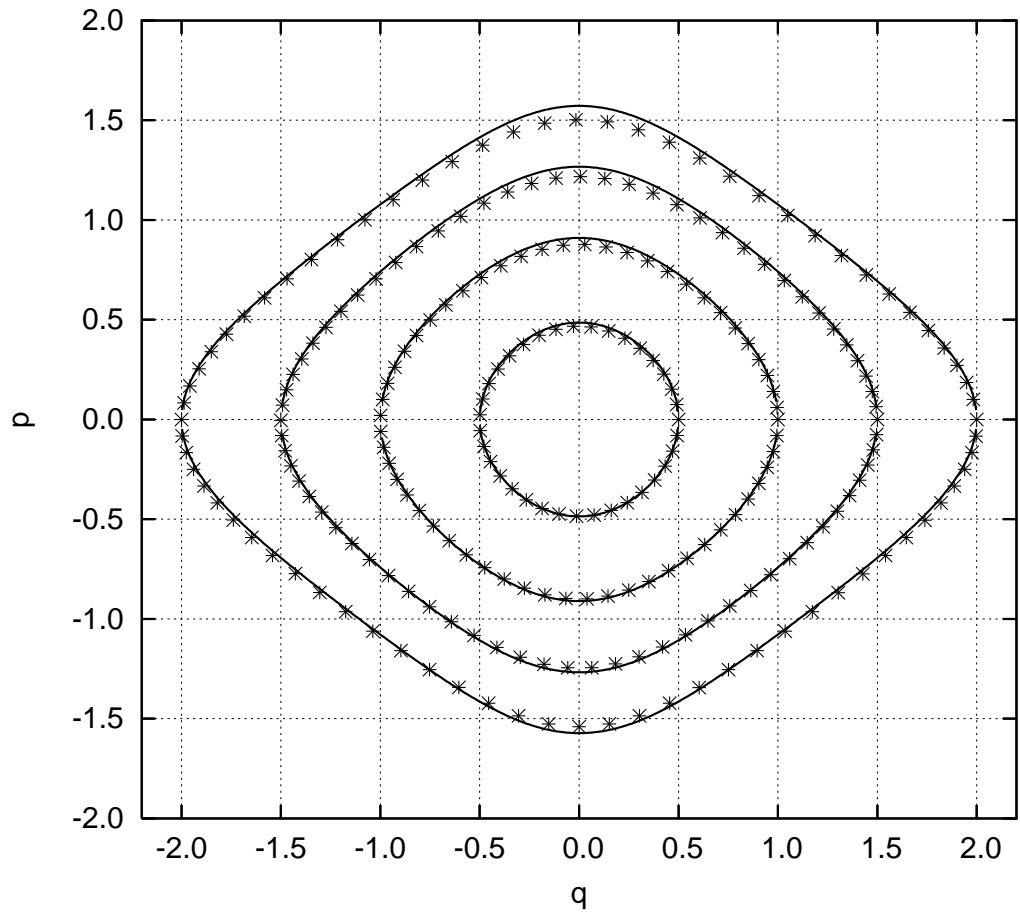}}
	\vspace{0.5truein}
\caption{ The phase trajectories of the non-separable Hamiltonian (\ref{h1d}).
The computed phase points (stars) are compared with exact trajectories (lines).
The initial values are $p_0=0$ and $q_0=0.5$, 1.0, 1.5 and 2.0, corresponding
to energy values of 1.25, 2.0, 3.25 and 5.0 respectively.
\label{fig1}}
\end{figure}
\newpage
\begin{figure}
	\vspace{0.5truein}
	\centerline{\includegraphics[width=0.8\linewidth]{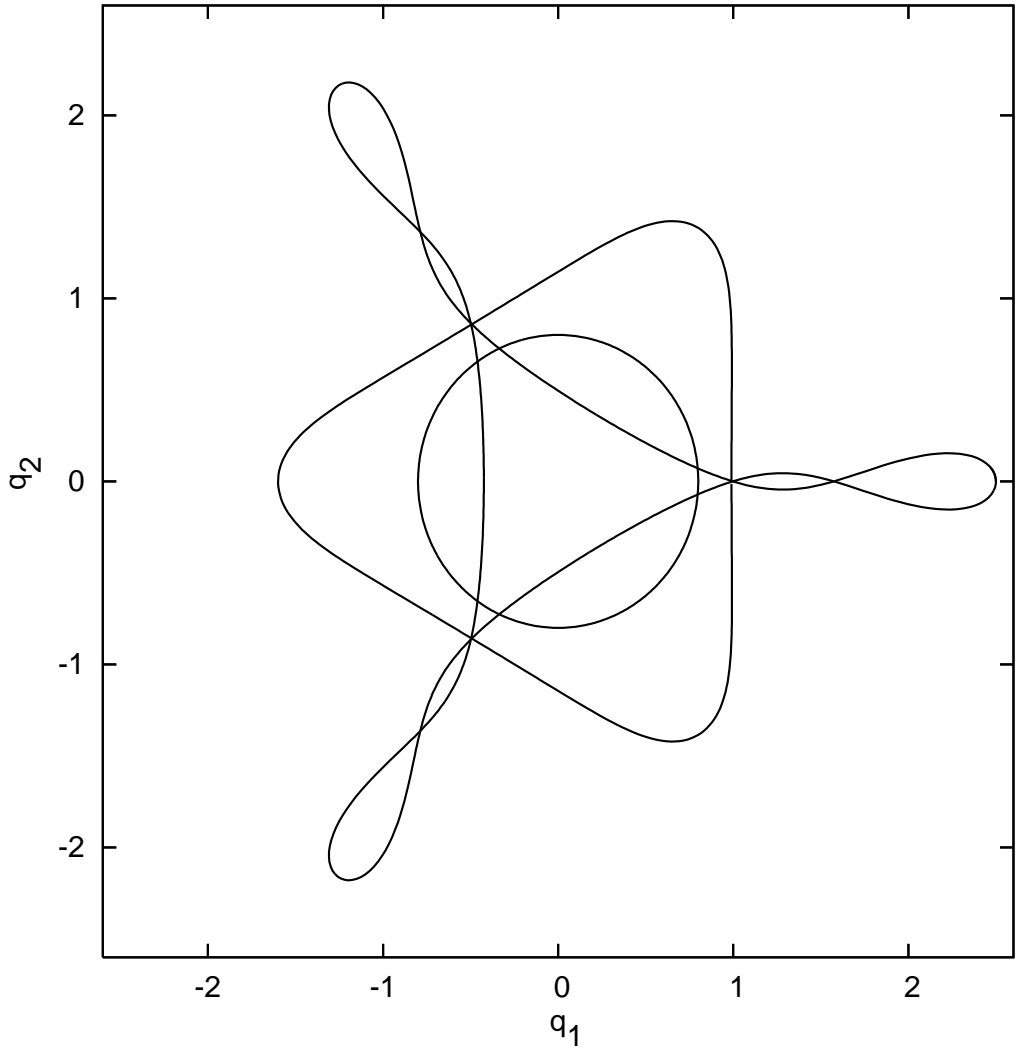}}
	\vspace{0.5truein}
\caption{ Some two dimensional orbits of the non-separable 
Hamiltonian (\ref{hnd}). Most trajectory are not closed and only very special 
initial conditions can result in closed orbits.
The initial conditions $(q_1,q_2,p_1,p_2)$ that produce the circle,
the triangle and the twisted orbits are respectively,
(0.8,0,0,0.425), (0.99,0,0,0.789) and (2.5,0,0,0.1884).  
\label{fig2}}
\end{figure}
\newpage												
\begin{figure}
	\vspace{0.5truein}
	\centerline{\includegraphics[width=0.8\linewidth]{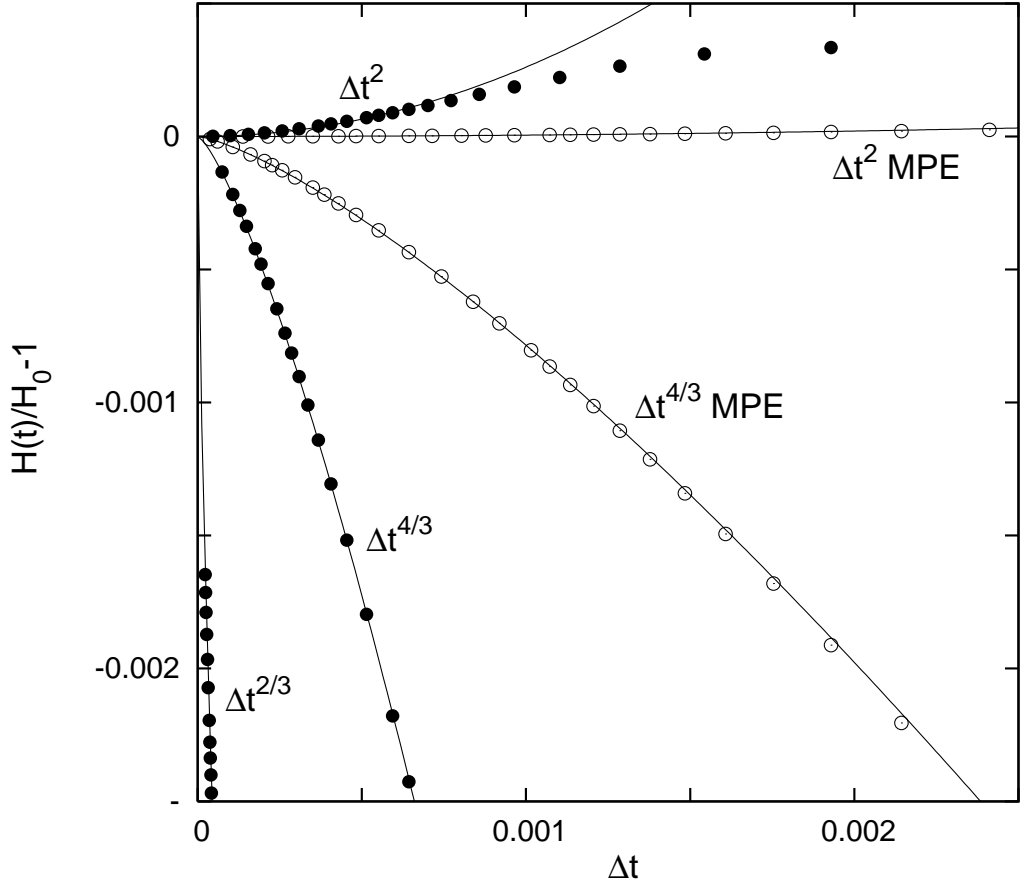}}
	\vspace{0.5truein}
\caption{The fractional power convergence of various explicit algorithms.
The relative energy error is evaluated at the first quarter period $t=0.385841$, 
for the outermost trajectory of Fig.1. The solid circles denote results of
symplectic algorithms (\ref{expalg}), (\ref{sym43}) and (\ref{sym63}). The hallow
circles give results of MPE algorithms (\ref{mpe43}) and (\ref{mpe63}). The lines
are fitted cuves of the form $c\dt^n$, with $n=2/3, 4/3,$ or 2 as indicated.
\label{fig3}}
\end{figure}
\end{document}